\documentclass[apj]{emulateapj}

\usepackage{graphicx,times}
\usepackage{subfigure}
\newcommand{\be}{\begin{equation}}
\usepackage{threeparttable}
\usepackage{booktabs}
\newcommand{\ee}{\end{equation}}
\newcommand{\bea}{\begin{eqnarray}}
\newcommand{\eea}{\end{eqnarray}}

\usepackage{enumerate}
\usepackage{amsmath}
\usepackage{cases}
\usepackage{longtable}
\usepackage{hyperref}
\usepackage{epstopdf}
\usepackage{amsmath,bm}
\usepackage{amssymb}
\usepackage{natbib}
\usepackage{morefloats}
\usepackage{multirow}
\usepackage{array}
\usepackage{verbatim}

\begin{document}

\title{Scatter broadening of pulsars and implications on the interstellar medium turbulence}

\author{Siyao Xu\altaffilmark{1} and Bing Zhang \altaffilmark{1,2,3}}

\altaffiltext{1}{Department of Astronomy, School of Physics, Peking University, Beijing 100871, China; syxu@pku.edu.cn}
\altaffiltext{2}{Kavli Institute for Astronomy and Astrophysics, Peking University, Beijing 100871, China}
\altaffiltext{3}{Department of Physics and Astronomy, University of Nevada Las Vegas, NV 89154, USA; zhang@physics.unlv.edu}

\begin{abstract}

Observations reveal a uniform Kolmogorov turbulence throughout the diffuse ionized interstellar medium (ISM)
and supersonic turbulence preferentially located in the Galactic plane.
Correspondingly, we consider the Galactic distribution of electron density fluctuations consisting of not only 
a Kolmogorov density spectrum but also a short-wave-dominated density spectrum with the density structure formed at small scales 
due to shocks. 
The resulting dependence of the scatter broadening time on the dispersion measure (DM) naturally interprets the existing observational data 
for both low and high-DM pulsars. 
According to the criteria that we derive for a quantitative determination of scattering regimes over wide ranges of DMs and frequencies $\nu$, 
we find that 
the pulsars with low DMs are primarily scattered by the Kolmogorov turbulence,
while those at low Galactic latitudes with high DMs undergo more enhanced scattering dominated by the supersonic turbulence, where the corresponding 
density spectrum has a spectral index $\approx 2.6$.
Besides, by considering a volume filling factor of the density structures with the dependence on $\nu$ as $\propto \nu^{1.4}$
in the supersonic turbulence, 
our model can also explain the observed shallower $\nu$ scaling of the scattering time than the Kolmogorov scaling for the pulsars with relatively large DMs. 
The comparison between our analytical results and the scattering measurements of pulsars in turn makes a useful probe
of the properties of the large-scale ISM turbulence, e.g., an injection scale of $\sim 100$ pc, 
and also characteristics of small-scale density structures.

\end{abstract}

\keywords{stars: pulsars: general - scattering  - turbulence - ISM: structure}

\section{Introduction}

The substantial population of Galactic pulsars enables sufficient sampling of the turbulent density in the ISM 
by many lines of sight (LOS) toward them.
Pulsar signals that traverse through the fluctuating density field undergo multi-path scattering, with the 
radio pulses broadened in time
\citep{Will72}.
The scatter broadening time has a strong dependence on both the interstellar dispersion and frequency 
\citep{Sch68, Ro86}.
Their scaling relations comply with the distribution of electron density fluctuations in the interstellar space.
Thus interstellar scattering measurements of pulsar radiation offer a valuable opportunity 
for statistical studies on the nature of ISM turbulence.

On the other hand, 
a clear physical interpretation of the observed pulse broadening phenomenon requires a good knowledge of the interstellar electron density structure. 
A power-law model of electron density fluctuations is commonly adopted in theoretical constructions on radio wave propagation
\citep{Lee76, Rick77, Ric90}
and compatible with observational indications 
(e.g. \citealt{Armstrong95}).
Recent advances in understanding the properties of magnetohydrodynamic (MHD) turbulence 
\citep{GS95, LG01, CL02_PRL,CL03}
stimulate a renewed investigation on density statistics
\citep{BLC05,KL07,Lad08,Burk09,Burk10, Col12,Fed12, Bur15},
which provide important insight into key physical processes such as star formation  
in the turbulent and magnetized ISM
(see reviews by e.g., \citealt{Mckee_Ostriker2007, Laz15}).
The density spectrum in compressible astrophysical fluids was systematically studied in 
\citet{KL07}
by carrying out an extensive set of MHD numerical simulations with varying compressibility and magnetization.
Instead of a single Kolmogorov slope with the power-law index of $\beta=11/3$, 
significant variations in the spectral slope of density fluctuations are present.
For supersonic turbulence, their results 
are consistent with earlier findings in both magnetized 
\citep{BLC05} and nonmagnetized  
\citep{Kim05}
fluids. It shows that 
the density power spectrum becomes shallower as the sonic Mach number ($M_s=V_L/c_s$) increases, where $V_L$ is the turbulent velocity 
at the outer scale of turbulence and $c_s$ is the sound speed in the medium, 
and there is a significant excess of density structures at small scales in highly supersonic turbulence.
This behavior is naturally expected as the gas is compressed in shocks by supersonic flows 
and the interacting shocks produce local density enhancements
\citep{PJ04, MacL04}.

The ISM exists in various phases with different physical properties
\citep{Spa01}. 
A number of new observational techniques on measuring $M_s$ in the turbulent ISM has been developed recently 
(see \citealt{Bla12} and references therein).
The warm ionized medium (WIM) is a major component of the diffuse and ionized ISM 
\citep{Hi08, Haf09},
and has a volume filling factor of $\sim 25 \%$
\citep{Tie,HS13}.
The estimated $M_s$ of the WIM is of order unity 
\citep{kul87, Haf99, Hi08}.
The statistical analysis of the gradient of linearly polarized radio emission also suggests that the turbulence in the WIM is 
subsonic to transonic
\citep{Gae11, Bur12}.
As expected for subsonic and transonic turbulence,
density fluctuations act as a passive scalar and follow the same Kolmogorov spectrum as turbulent velocity,
which spans from $10^{-5}$ AU up to an inferred outer scale on the order of $100$ pc
and is known as the ``big power law in the sky"
\citep{Armstrong95, CL10}.
Such a large injection scale of turbulence was also reported in 
\citet{Hav06, Hav08}
by measuring structure functions of Faraday rotation measure for Galactic interarm regions, suggesting the main sources of turbulence in the WIM 
as supernova and superbubble explosions 
(see review by \citealt{HS13}). 
In other colder and denser phases in the inner Galactic plane, such as the cold neutral medium and molecular clouds,
the turbulence is supersonic with $M_s >1$
(e.g., $M_s \approx 5-20$ in molecular clouds, see 
\citealt{Zuc74, Lars81}),
and consists of a network of shocks.
Density fluctuations and velocity fluctuations exhibit distinct power spectra
\citep{Fal14}.
\footnote{Unlike the density spectrum which can have the spectral index either higher or lower than $\beta=3$, 
turbulent velocity spectrum always has $\beta>3$
\citep{CL03v},
and it becomes even steeper than the Kolmogorov scaling in supersonic turbulence
(see simulations by, e.g. \citealt{Kri07,Sch09,Fed10,KL10}
and observations by, e.g. \citealt{Pa06,Pa09,CLS10}).}
The inference of very shallow spectra of density can be drawn from 
21~cm line absorption measurements
\citep{Des00},
and CO line emission of molecular clouds 
\citep{Stu98, Padoan04, Sw06}.
An ensemble of indices of density spectra lower than $\beta = 3$
that are extracted from spectroscopic data 
can be found in 
reviews by, e.g., 
\citet{Laz09rev,HF12}.
In addition, in comparison with the subsonic to transonic turbulence in the diffuse WIM, 
the supersonic turbulence that resides in the Galactic plane may have a small outer scale of a few parsecs 
associated with the stellar source of turbulent energy
\citep{Hav08, Malk10}
and not contribute to large-scale density fluctuations in the Galaxy.
Moreover, within the cold and dense ISM phases which have a small filling factor ($\sim 1\%$ for the cold neutral medium
and $\sim 0.05\%$ for molecular clouds, \citealt{Tie,HS13}), 
the supersonic turbulence creates considerably high density contrasts and 
small-scale density structures with a further smaller filling factor.

In accordance with the distinctive turbulence properties in different ISM phases, 
the Galactic distribution of electron density fluctuations is expected to consist of 
a Kolmogorov density spectrum in the subsonic to transonic turbulence throughout the diffuse ionized ISM as shown by the 
``big power law in the sky"
\citep{Armstrong95, CL10},
and a shallower density spectrum with $\beta<3$ in the supersonic turbulence prevalent in the inner Galactic plane. 
Regarding the latter case, 
despite the ample measurements of neutrals (see above),
to our knowledge, such a shallow density spectrum of electron density fluctuations has only been extracted from the rotation measurements of polarized extragalactic sources 
\citep{XuZ16}.
Potentially, the scattering measurements of low-latitude pulsars enable us to carry out a more detailed 
investigation of the electron density distribution in the supersonic turbulence in the Galactic plane.
Conventionally, it is the canonical Kolmogorov distribution of density irregularities that has been adopted in 
early attempts to understand the pulsar scattering observations and properties of the interstellar turbulence
\citep{Lee76, Arm81, Armstrong95, CoR98, Cor16}.
Indeed, the observed scalings of pulse broadening time with both DM 
\citep{Ram97,Loh04,Kri15}
and frequency 
\citep{Cor85, Joh98, Loh01,Loh04, Lew13}
for low-DM pulsars (DM$<100$ pc cm$^{-3}$) are in agreement with the Kolmogorov's theory predictions, irrespective of Galactic latitudes.
On the other hand, significant deviations from the Kolmogorov scaling are commonly seen in scatter broadening measurements of high-DM pulsars
\citep{Loh01, Loh04, Bha04, Lew13, Lew15, Kri15, Cor16}.
These discrepant observations eliminate a single power law for a global description of density spectra within the ISM. 
Besides a homogeneous component corresponding to the Kolmogorov turbulence, 
a clumped medium with discrete clumps and voids has been suggested 
to account for the inhomogeneity of the ISM, e.g., the variation of scattering strength with path length and Galactic latitude 
\citep{Cor85}, 
and to model the Galactic distribution of free electrons, e.g., the NE2001 model 
\citep{Cor02, NE2}.

Motivated by both the numerical and observational evidence, 
we consider a spectral model for interstellar electron density fluctuations by incorporating not only a Kolmogorov density spectrum with 
$\beta=11/3$ but also a shallower density spectrum with $\beta<3$,
to perform a comprehensive analysis of the interstellar scattering of pulsars.
Moreover, the second-order density statistics in a turbulent flow, namely, 
the density spectrum in Fourier space and the structure function of density fluctuations in real space
\citep{LP04, LP06},
can be used to recover statistical properties of the ISM turbulence, 
which imprint on observables such as 
velocity centroids
\citep{LE03,EL05,Bur14},
Doppler-shifted emission and absorption spectral lines
\citep{LP00, LP04, LP06, Las06},
rotation measure fluctuations 
\citep{MS96, XuZ16},
as well as the scatter broadening time of pulsars that we focus on in the current study. 
The observationally measured DM and frequency scalings of the pulse broadening time impose constraints on the 
slope, amplitude, and cutoff scales of the density power spectrum. 
This can provide information on the injection and transfer of turbulent energy in the WIM  
where the density can be treated as a passive scalar transported by the turbulent velocity field, 
and on the small-scale density structures in highly supersonic turbulence in the inner Galaxy.

In Section 2, we present a general formalism for the scalings of scattering time with DM and observing frequency 
for a power-law spectrum of electron density fluctuations.
In Section 3, by comparing the analytical results with the scatter broadening measurements of pulsars, 
we model the distribution of interstellar density fluctuations and identify the scattering regimes over 
different ranges of DMs and frequencies. 
The discussion and conclusions are given in Section 4 and 5.

\section{Scalings of scattering time with DM and frequency}\label{sec: scal}

We consider a power-law spectrum of electron density fluctuations with the outer and inner scales of the turbulence as
$L$ and $l_0$
\citep{Rick77, Col87},
\begin{equation}\label{eq: orisf}
    P(k) = C_N^2 k^{-\beta} e^{-(kl_0)^2}, ~~ k> L^{-1}, 
\end{equation}
where the spectral index $\beta$ of the 3D power spectrum is within the range $2<\beta<4$
\footnote{The density spectrum in the interstellar turbulence steeper than $\beta=4$ is rejected 
since its associated refractive modulation index is inconsistently larger than that observed from the nearby pulsars 
\citep{RicL90, Armstrong95, Lam00}.}.
The density spectrum with $\beta>3$
is termed {\it a long-wave-dominated density spectrum} and characterized by large-scale density fluctuations, 
while {\it a short-wave-dominated density spectrum} refers to the density spectrum with $\beta<3$ and describes small-scale density structures
\citep{LP00,LP04,LP06}.
The coefficient $C_N^2$ represents the scattering strength per unit length along the LOS. 
It is determined by the root-mean-square (rms) amplitude of density fluctuations $\delta n_e$ at the density correlation scale, 
which is $L$ for a long-wave-dominated density spectrum and 
$l_0$ for a short-wave-dominated density spectrum
\citep{LP16},
\begin{subnumcases}
    {C_N^2 \sim \label{eq: cnsss}}
    \mathcal{C}(\beta) (\delta n_e)^2 L^{3-\beta},   ~~~ \beta > 3, \label{eq: cnstep} \\
    \mathcal{C}(\beta) (\delta n_e)^2 l_0^{3-\beta}, ~~~~ \beta < 3, \label{eq: cnshal}
\end{subnumcases}
where 
\begin{subnumcases}
  {\mathcal{C}(\beta) = \label{eq: betpa}}
   \frac{\beta-3} {2(2\pi)^{4-\beta}},  ~~~ \beta > 3, \\   
   \frac{3-\beta}{2(2\pi)^{4-\beta}}, ~~~ \beta < 3.
\end{subnumcases}
The path integral of $C_N^2$ along the LOS to the pulsar at a distance $D$ is the scattering measure SM
\citep{Cor02, NE2}, 
which for a LOS through a statistically uniform scattering medium is simplified as 
\begin{equation}\label{eq: simsm}
     \text{SM}  =  C_N^2 D.
\end{equation}
Radio wave scattering by a turbulent medium introduces phase fluctuations to the wavefront. 
Corresponding to the density power spectrum given by Eq. \eqref{eq: orisf}, 
the phase structure function $D_\Phi$ under the consideration of $r \ll L \ll D$ is 
\citep{Col87,Ric90}
\begin{subnumcases}
    {D_\Phi = \label{eq: sfspec}}
     \pi r_e^2 \lambda^2 \text{SM} l_0^{\beta-2} \Big(\frac{r}{l_0}\Big)^2,  ~~ r<l_0, \label{eq: dsfcf}\\
     \pi r_e^2 \lambda^2 \text{SM} r^{\beta-2}, ~~~~~~~~~~~~ r>l_0, \label{eq: dsfiner}
\end{subnumcases}
where $r_e$ is the classical electron radius, $\lambda$ is the wavelength, and $r$ is the transverse spatial separation between a pair of LOSs.
The transverse scale over which the rms phase difference is 1 radian, i.e., $D_\Phi =1$,
is defined as the diffractive scale $r_\text{diff}$.
We next discuss the cases of $r_\text{diff}<l_0$ and $r_\text{diff}>l_0$, respectively.

(1) $r_\text{diff}<l_0$~~~~
In a particular case of $r=l_0=r_\text{diff}$, 
by inserting Eq. \eqref{eq: cnsss} and \eqref{eq: simsm} in Eq. \eqref{eq: sfspec}, 
we find 
\begin{equation}\label{eq: parcad}
   D_\Phi = \pi r_e^2 \Gamma(\beta) \lambda^2 D =1.
\end{equation}
Here all the quantities related to the spectral properties of turbulent density are contained in the function 
\begin{equation}
   \Gamma(\beta) = \mathcal{C}(\beta) (\delta n_e (l_0))^2 l_0,
\end{equation}
where the density perturbation at $l_0$ is given according to the power-law scaling,
\begin{subnumcases}
    {(\delta n_e(l_0))^2 =\label{eq: denss}}
     (\delta n_e)^2 \Big(\frac{l_0}{L}\Big)^{\beta-3},   ~~ \beta > 3, \\
     (\delta n_e)^2, ~~~~~~~~~~~~~~~~~ \beta < 3.
\end{subnumcases}
Then in the case when $r_\text{diff}$ is below $l_0$, one expects $D_\Phi (r=l_0)>1$, by using expression in Eq. \eqref{eq: parcad}, 
which requires 
\begin{equation}\label{eq: innerdm}
   \text{DM} > [\pi r_e^2 \Gamma(\beta) \lambda^2]^{-1} n_e, 
\end{equation}
at a given $\lambda$. The dispersion measure of the scattering medium 
is defined as DM $= n_e D$, where $n_e$ is the LOS average electron density. 
For an individual source with a fixed DM, the condition $D_\Phi (r=l_0)>1$ is satisfied with 
\begin{equation}\label{eq: innernu}
   \nu < \Big[\frac{1}{n_e}\pi r_e^2 \Gamma(\beta) c^2 \text{DM}\Big]^\frac{1}{2},
\end{equation}
where $\nu = c/ \lambda$ is the frequency. Eq. \eqref{eq: innerdm} and \eqref{eq: innernu} indicate the 
ranges of DM and $\nu$ where the effect of the inner scale of density spectrum on scattering of pulsar signals should be considered.

From the condition $D_\Phi =1$ and Eq. \eqref{eq: dsfcf}, $r_\text{diff}$ in the case of $r_\text{diff}<l_0$ has the expression
\begin{equation}\label{eq: rdfsm}
   r_\text{diff} = \Big(\pi r_e^2 \lambda^2 \text{SM} l_0^{\beta-4}\Big)^{-\frac{1}{2}}.
\end{equation}
With the parameters absorbed into $r_\text{diff}$, 
$D_\Phi$ from Eq. \eqref{eq: sfspec} can be conveniently written as 
\begin{subnumcases}
 {D_\Phi =}
 \Big(\frac{r}{r_\text{diff}}\Big)^2,  ~~~~~~~~~~~~~~~~~~~~~~ r < l_0, \\
 \Big(\frac{l_0}{r_\text{diff}}\Big)^2 \Big(\frac{r}{l_0}\Big)^{\beta-2}, ~~~~~~~r>l_0,
\end{subnumcases}
which is a broken power-law with a shallower slope on scales larger than $l_0$.

The scattering observable of interest is the scatter broadening time, which is related to $r_\text{diff}$ by  
\begin{equation}\label{eq: sctime}
   \tau_\text{sc} =  \frac{D \lambda^2}{ 4 \pi^2 c }  r_\text{diff}^{-2}.
\end{equation}
By inserting Eq. \eqref{eq: cnsss}, \eqref{eq: simsm}, \eqref{eq: denss}, and \eqref{eq: rdfsm} in the above equation, $\tau_\text{sc}$ has the form
\begin{equation}\label{eq: dissdm}
   \tau_\text{sc} =  \frac{r_e^2 c^3}{4\pi} \mathcal{C}(\beta) \Big(\frac{\delta n_e (l_0)}{n_e}\Big)^2 l_0^{-1} \text{DM}^2 \nu^{-4}.
\end{equation}
From the above expression we can write the dependence of $\tau_\text{sc}$ on DM and $\nu$ as 
\begin{equation}\label{eq: innscal}
     \tau_\text{sc} \propto   \text{DM}^\alpha \nu^{-2\alpha}, ~~\alpha = 2.
\end{equation}

We notice that in the case of $r_\text{diff}<l_0$, 
the Gaussian form of density distribution on scales smaller than the inner cutoff $l_0$ of the density power spectrum 
(see Eq. \eqref{eq: orisf}) 
leads to the same result as a Gaussian distribution of density irregularities 
\citep{Sch68, La71,Lee76}
with the fluctuating electron density $\delta n_e (l_0)$ and characteristic scale $l_0$.
The strong scattering is dominated by the density perturbation at $l_0$, and the resulting DM and frequency scalings have  
a critical minimum value of $\alpha$
\citep{Ro86}.

(2) $r_\text{diff}>l_0$~~~~
When $r_\text{diff}$ exceeds $l_0$, 
there is $D_\Phi (r=l_0)<1$, 
which sets the upper limit of DM at a given $\nu$
\begin{equation}
   \text{DM} < [\pi r_e^2 \Gamma(\beta) \lambda^2]^{-1} n_e, 
\end{equation}
and the lower limit of $\nu$ at a given DM
\begin{equation}
   \nu > \Big[\frac{1}{n_e}\pi r_e^2 \Gamma(\beta) c^2 \text{DM}\Big]^\frac{1}{2}.
\end{equation}
The diffractive scale calculated from $D_\Phi =1$ by using Eq. \eqref{eq: dsfiner} for $r_\text{diff}>l_0$ is
\begin{equation} \label{eq: rdiff}
   r_\text{diff} =  \Big(\pi r_e^2 \lambda^2 \text{SM}\Big)^\frac{1}{2-\beta}.
\end{equation}
Substitution of the above expression into Eq. \eqref{eq: sfspec} gives 
\begin{subnumcases}
 {D_\Phi =}
 \Big(\frac{l_0}{r_\text{diff}}\Big)^{\beta-2} \Big(\frac{r}{l_0}\Big)^2, ~~~~~~~~~~~ r < l_0, \\
 \Big(\frac{r}{r_\text{diff}}\Big)^{\beta-2}, ~~~~~~~~~~~~~~~~~~~~~~r>l_0.
\end{subnumcases}

The scattering time can be obtained by inserting Eq. \eqref{eq: cnsss}, \eqref{eq: simsm}, and \eqref{eq: rdiff} into Eq. \eqref{eq: sctime},  
\begin{subnumcases}
    {\tau_\text{sc} = \label{eq: inerdm}}
 \mathcal{F} L^\frac{2(3-\beta)}{\beta-2} \text{DM}^\frac{\beta}{\beta-2} \nu^{-\frac{2\beta}{\beta-2}}, ~~~~~~~~ \beta>3,   \label{eq: wdsstb} \\
 \mathcal{F} l_0^\frac{2(3-\beta)}{\beta-2} \text{DM}^\frac{\beta}{\beta-2} \nu^{-\frac{2\beta}{\beta-2}},  
  ~~~~~~~~~ \beta<3,   \label{eq: wdsshb}
\end{subnumcases}
where 
\begin{equation}
     \mathcal{F}= \frac{r_e^\frac{4}{\beta-2}  c^\frac{\beta+2}{\beta-2}  }{4\pi^\frac{2(\beta-3)}{\beta-2}    } 
           \mathcal{C}(\beta)^\frac{2}{\beta-2} \Big(\frac{\delta n_e}{n_e}\Big)^\frac{\beta}{\beta-2} (\delta n_e)^\frac{4-\beta}{\beta-2}.
\end{equation}
It shows that $\tau_\text{sc}$ can also be expressed in terms of DM and $\nu$ with the same form as in Eq. \eqref{eq: innscal}
\begin{equation}\label{eq: screst}
     \tau_\text{sc} \propto   \text{DM}^\alpha \nu^{-2\alpha}, 
\end{equation}
but instead of a constant, here $\alpha$ is related to the spectral index by 
\begin{equation}
  \alpha = \beta/(\beta-2), 
\end{equation}
and falls in different ranges for long- and 
short-wave-dominated density spectra, 
\begin{equation}
\begin{aligned}
   & 2< \alpha <3, ~~~~~~~~~ 3<\beta<4, \\
   & 3< \alpha < + \infty, ~~~~ 2<\beta<3.  
\end{aligned}
\end{equation}
Notice that for the long-wave-dominated Kolmogorov density spectrum with $\beta=11/3$, the corresponding value of $\alpha$ is $2.2$.

From both scaling relations presented in Eq. \eqref{eq: innscal} and \eqref{eq: screst}, we see that 
the scattering timescale decreases with $\nu$, showing more pronounced scattering at lower frequencies. 
Meanwhile, it increases with DM, which is an indicator of the distance,
i.e., the thickness of the turbulent scattering medium between the pulsar and the observer. 
In comparison with the case of $r_\text{diff}<l_0$, 
evidently, when $r_\text{diff}>l_0$, 
$\tau_\text{sc}$ has a stronger dependence on DM, 
and the trend steepens with decreasing $\beta$, indicative of stronger scattering toward higher DMs 
for a shallower density spectrum.

\section{Application to scatter broadening measurements of pulsars}

\subsection{The spectral model for Galactic distribution of electron density fluctuations}

The electron density spectrum throughout the diffuse WIM
has been demonstrated to comply with the well-known Kolmogorov power law 
\citep{Armstrong95, CL10}.
Accordingly, we adopt the Kolmogorov model with $\beta=11/3$ for the homogeneous component of the interstellar turbulent density field,
which serves as a uniformly pervasive scattering medium in the ISM. 
The scattering time deduced from the Kolmogorov scattering statistics is (Eq. \eqref{eq: wdsstb})
\begin{equation}\label{eq: modste}
\begin{aligned}
    \tau_\text{sc} =  1.2\times10^4
     & \Big(\frac{\delta n_e}{n_e}\Big)^{2.2} \Big(\frac{\delta n_e}{0.01 \text{cm}^{-3}}\Big)^{0.2} 
\Big(\frac{L}{100 \text{pc}}\Big)^{-0.8} \\
                                                           &  \Big(\frac{\text{DM}}{ \text{pc cm}^{-3}}\Big)^{2.2} \Big(\frac{\nu}{ \text{MHz}}\Big)^{-4.4} \text{ms}.
\end{aligned}
\end{equation}
Positive evidence for the above scaling $ \tau_\text{sc} \propto \text{DM}^{2.2} \nu^{-4.4}$ 
\citep{Ro86}
can be found from observations of relatively nearby pulsars at both high and low Galactic latitudes
\citep{Cor85, Joh98, Stin00, Lam00, Lew13}.

On the other hand, there are substantial observational inconsistencies with the Kolmogorov density spectrum indicated from the 
scattering measurements of high-DM pulsars
(e.g., \citealt{Loh01, Loh04, Bha04}).
To produce the more enhanced scattering observed for the distant pulsars at low Galactic latitudes
\citep{Cor16},
a flatter density spectrum with larger density fluctuations on small scales in comparison with the Kolmogorov spectrum are required
for modeling the distribution of turbulent density in the inner Galaxy.
Such a short-wave-dominated density spectrum is confirmed by 
numerical simulations of compressible turbulence with a high sonic Mach number
\citep{Kim05, BLC05, KL07}
and observed toward the inner Galactic plane where the density field is highly structured as a result of shock compressions in 
supersonic turbulence
\citep{Laz09rev,HF12}.
The resulting small-scale overdense structures generated in the cold and dense ISM phases with a small filling factor can only 
occupy a further smaller fraction of the volume that the LOS passes through. Accordingly, 
a volume filling factor $f$ that is much less than unity 
needs to be included when applying a short-wave-dominated density spectrum
to quantify the strengthened scattering effect. 
By replacing the rms density perturbation $\delta n_e$ with $\sqrt{f} \delta n_e$ in Eq. \eqref{eq: wdsshb}, we have $\tau_\text{sc}$ as
\begin{equation}\label{eq: modsha}
\begin{aligned}
   \tau_\text{sc} =  &
 \frac{ r_e^\frac{4}{\beta-2}  c^\frac{\beta+2}{\beta-2} }{4 \pi^\frac{2(\beta-3)}{\beta-2}}   
 \mathcal{C}(\beta)^\frac{2}{\beta-2}  \Big(\frac{\sqrt{f}\delta n_e}{n_e}\Big)^\frac{\beta}{\beta-2} (\sqrt{f}\delta n_e)^\frac{4-\beta}{\beta-2}
 l_0^\frac{2(3-\beta)}{\beta-2}  \\
                                   &  \text{DM}^\frac{\beta}{\beta-2} \nu^{-\frac{2\beta}{\beta-2}} .
\end{aligned}
\end{equation}

An inverse correlation between $f$ and the average density of a density structure in the diffuse ionized ISM 
has been indicated in earlier theoretical  
\citep{Fle96, Elm99}
and numerical 
\citep{El97,Ko07}
studies, 
as well as in observations 
\citep{Gau93, Ber06, Ber08}.
Moreover, observations also suggest that the correlation becomes considerably steeper at low latitudes
than in the diffuse ionized gas away from the Galactic plane 
\citep{Ber08}.
Based on both theoretical and observational grounds, we assume that $f$ and the density fluctuation $\delta n_e$ is anti-correlated. 
According to the power-law scaling of the density spectrum, $\delta n_e$ increases  
toward smaller scales for a short-wave-dominated density spectrum. 
Therefore, smaller-scale density structures possess a smaller $f$.
Meanwhile, the diffractive scattering of lower-frequency waves is mainly attributed to the density fluctuations at smaller scales 
(Eq. \eqref{eq: rdiff}). 
In view of the above arguments, we consider a $\nu$-dependent $f$, 
\begin{equation}\label{eq: defres}
    f = f_0 \Big(\frac{\nu}{\nu_0}\Big)^\eta, ~~\eta>0,
\end{equation}
where $f_0$ is the filling factor corresponding to the reference frequency $\nu_0$. 
By inserting the above expression into Eq. \eqref{eq: modsha}, we derive 
\begin{equation}\label{eq: nudeff}
\begin{aligned}
   \tau_\text{sc} =  &
   \frac{ r_e^\frac{4}{\beta-2}  c^\frac{\beta+2}{\beta-2} }{4 \pi^\frac{2(\beta-3)}{\beta-2}}   
    (\mathcal{C}(\beta) g_0)^\frac{2}{\beta-2}  \Big(\frac{\delta n_e}{n_e}\Big)^\frac{\beta}{\beta-2} (\delta n_e)^\frac{4-\beta}{\beta-2}
    l_0^\frac{2(3-\beta)}{\beta-2}  \\
                                   &  \text{DM}^\frac{\beta}{\beta-2} \nu^{\frac{2(\eta-\beta)}{\beta-2}} ,
\end{aligned}
\end{equation}
where the constant $f_0 \nu_0^{-\eta}$ is replaced by the factor $g_0$. 
By taking into account
the dependence of $f$ on $\nu$, the scaling of $\tau_\text{sc}$ with $\nu$ is modified. 
The values of $\beta$ and $\eta$ depend on the compressibility and magnetization of the scattering medium. 
A comparison with the temporal broadening measurements of heavily scattered sources can provide constraints on the 
actual spectral form and $f$ of density structures.

Therefore, we consider 
(a) a highly structured density field with the excess of density fluctuations over small scales described by a short-wave-dominated 
density spectrum,
\footnote{Visualizations of the density structures developed in simulations of supersonic turbulence show the prevalence of filaments and sheets
\citep{Kim05, KL07}.} 
embedded in 
(b) a uniformly distributed turbulent medium with a Kolmogorov density spectrum, 
corresponding to the distribution of electron density fluctuations in the supersonic and Kolmogorov turbulence, respectively. 
In the case of the supersonic turbulence, we carry out the calculations by adopting both a constant $f$ as a simplified approach (Eq. \eqref{eq: modsha}), 
and a $\nu$-dependent $f$ as a more realistic treatment (Eq. \eqref{eq: nudeff}).
We next confront this model for the distribution of interstellar density fluctuations 
with the scattering measurements of pulsars.

\subsection{Comparison with pulsar observations}\label{ssec: com}

\citet{Kri15} 
presented the $\tau_\text{sc}$ measurements for 124 pulsars at $327$ MHz, including some known samples collected from the literature.
Earlier measurements at different frequencies were all referenced to $\nu = 327$ MHz by using 
the Kolmogorov frequency scaling $\tau_\text{sc} \propto \nu^{-4.4}$.
The best fit to their data takes the form 
\citep{Kri15} 
\begin{equation}\label{eq: comfm}
\begin{aligned}
    \tau_\text{sc}  = & 3.6 \times 10^{-6} \Big(\frac{\text{DM}}{\text{pc cm}^{-3}}\Big)^{2.2}   \\
                                                        &  \Big[1 + 1.94\times 10^{-3} \Big(\frac{\text{DM}}{\text{pc cm}^{-3}}\Big)^{2.0} \Big] \text{ms},
\end{aligned}
\end{equation}
which corresponds to the empirical relation for scattering proposed by 
\citet{Ram97},
\begin{equation}
    \tau_\text{sc}  = A~ \text{DM}^\gamma (1+ B~\text{DM}^\zeta).
\end{equation}
In the low-DM range, the fit is dominated by 
\begin{equation}\label{eq: tsc1}
    \tau_\text{sc, low DM}  = 3.6 \times 10^{-6} \Big(\frac{\text{DM}}{\text{pc cm}^{-3}}\Big)^{2.2}  \text{ms}.
\end{equation}
The DM scaling is in accord with the Kolmogorov prediction for an insufficient scattering regime.
By comparing Eq. \eqref{eq: modste} with Eq. \eqref{eq: tsc1} and choosing suitable parameters for the homogeneous Kolmogorov turbulence, 
we derive 
\begin{equation}\label{eq: satconste}
\Big(\frac{n_e}{0.01 \text{cm}^{-3}}\Big)^{-2.2} \Big(\frac{\delta n_e}{0.04 \text{cm}^{-3}}\Big)^{2.4}
\Big(\frac{L}{100 \text{pc}}\Big)^{-0.8} \approx 1.2.
\end{equation}
The typical LOS average electron density $n_e$ is within the range $0.01-0.1$ cm$^{-3}$, but can have significant 
sightline-to-sightline variance 
\citep{NE2}.
The density perturbation $\delta n_e$ over the large turbulence injection scale is likely to be comparable to $n_e$. 
Under this consideration, the relation in Eq. \eqref{eq: satconste} suggests 
an outer scale $L$ of the Kolmogorov density spectrum comparable to the value ($\sim 100$ pc)
inferred from the observations of 
interstellar scattering of nearby pulsars 
\citep{Armstrong95}
and H$\alpha$ integrated intensity data for high Galactic latitudes
\citep{CL10}.
It is worthwhile to note that the outer scale value and the driving mechanism of the interstellar turbulence are still controversial. There is observational 
evidence showing the driving scale of turbulence on the order of kpc for some external galaxies
(e.g. \citealt{Che15}).
The relation shown in Eq. \eqref{eq: satconste} provides the constraint 
that the interstellar Kolmogorov turbulence should satisfy, so as to account for the degree of scattering for low-DM pulsars.

\begin{figure}[htbp]
\centering
   \includegraphics[width=8cm]{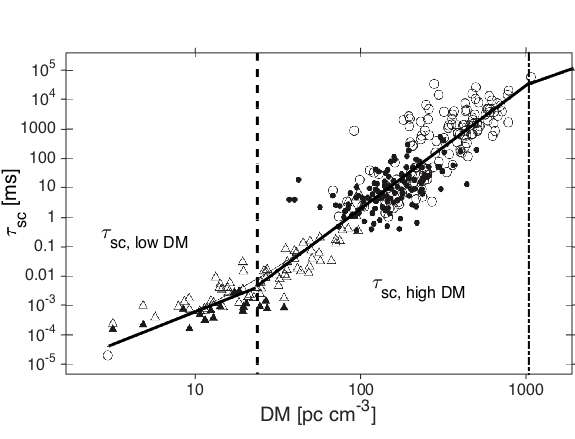}
\caption{The plot of $\tau_\text{sc}$ vs. DM at $327$ MHz taken from 
\citet{Kri15}.
The superposed thick solid lines represent $\tau_\text{sc, low DM}$ (Eq. \eqref{eq: tsc1}) and $\tau_\text{sc, high DM}$ (Eq. \eqref{eq: tsc2}) 
in low and high DM ranges, respectively. The vertical dashed line denotes $\text{DM}_\text{tr}$ (Eq. \eqref{eq: tran}). 
The vertical dash-dotted line corresponds to the relation in Eq. \eqref{eq: detreg} (or equivalently Eq. \eqref{eq: crsupff}) at $\nu = 327$ MHz, 
beyond which over further larger DMs, 
the $\tau_\text{sc}$-DM relation is given by Eq. \eqref{eq: shinn} (or equivalently Eq. \eqref{eq: shfrdf})
at $\nu = 327$ MHz. }
\label{fig: dataset}
\end{figure}

The fit at large DMs can be approximated by the functional form
\begin{equation}\label{eq: tsc2}
    \tau_\text{sc, high DM}  = 6.98\times10^{-9}  \Big(\frac{\text{DM}}{\text{pc cm}^{-3}}\Big)^{4.2} \text{ms}.
\end{equation}
The DM dependence of $\tau_\text{sc}$ is much stronger than the Kolmogorov theory expectation, and can only be explained by a 
short-wave-dominated density spectrum. 
By comparing Eq. \eqref{eq: modsha} and \eqref{eq: tsc2}, the spectral index is 
\begin{equation}\label{eq: avsslo}
   \frac{\beta}{\beta-2} = 4.2,  ~~\beta \approx 2.6. 
\end{equation}
By assuming a constant $f$, we are able to write Eq. \eqref{eq: modsha} as 
\begin{equation}
\begin{aligned}
      \tau_\text{sc} = &  5.0\times10^{-9} 
      \Big(\frac{f}{10^{-6}}\Big)^{3.2}  \Big(\frac{n_e}{0.01 \text{cm}^{-3}}\Big)^{-4.2} \\   
  &    \Big(\frac{\delta n_e}{0.1 \text{cm}^{-3}}\Big)^{6.4} 
         \Big(\frac{l_0}{10^7 \text{cm}}\Big)^{1.2}   \Big(\frac{\text{DM}}{\text{pc cm}^{-3}}\Big)^{4.2} \text{ms},
\end{aligned}
\end{equation}
for which to be consistent with Eq. \eqref{eq: tsc2}, there is 
\begin{equation}\label{eq: cntsha}
\begin{aligned}
 &  \Big(\frac{f}{10^{-6}}\Big)^{3.2}  \Big(\frac{n_e}{0.01 \text{cm}^{-3}}\Big)^{-4.2} \Big(\frac{\delta n_e}{0.1 \text{cm}^{-3}}\Big)^{6.4} 
   \Big(\frac{l_0}{10^7 \text{cm}}\Big)^{1.2}  \\
 &  \approx 1.4.
\end{aligned}
\end{equation}
It shows that the observed $\tau_\text{sc}$ trend at large DMs can be accommodated by a turbulent scattering medium which is 
characterized by a short-wave-dominated density spectrum with the spectral slope $\beta \approx 2.6$ (Eq. \eqref{eq: avsslo}) and turbulence parameters as indicated 
in the above equation. 
This spectral index derived from the scattering measurements of large-DM pulsars 
agrees well with the density spectral indices obtained from CO and 
H\uppercase\expandafter{\romannumeral1} in absorption
\citep{HF12}, 
and suggests $M_s \approx 7$ of the supersonic turbulent media 
by comparing with the numerical results in 
\citet{KL07,Burk10}.
In the case of a short-wave-dominated density spectrum, $\delta n_e$ for localized density enhancements on the inner scale $l_0$ 
is supposed to be noticeably
larger than the background density $n_e$. 
Besides, the turbulence inner scale  
in the ionized gas can be very small 
\citep{Spa90, Armstrong95}.
The filling factor $f$ required to match the data is smaller than unity by orders of magnitude, 
suggesting that sparsely distributed small-scale density structures 
are adequate to induce the intense scattering for high-DM pulsars.

When considering the $\nu$-dependent $f$, 
we first need multi-frequency measurements of $\tau_\text{sc}$ to determine the value of $\eta$ in Eq. \eqref{eq: nudeff}.
Departures from the Kolmogorov prediction and a flattening of the $\tau_\text{sc}$ spectrum with the slope $\leq 4$
for low-latitude and high-DM pulsars have been clearly shown by observations
\citep{Loh01, Loh04, Bha04, Lew13,Lew15}.
By adopting the scaling $\tau_\text{sc} \propto \nu^{-4}$ and using the $\beta$ value from Eq. \eqref{eq: avsslo}, we obtain 
\begin{equation}\label{eq: gamval}
    \frac{2(\eta-\beta)}{\beta-2} = -4, ~~ \eta \approx 1.4.
\end{equation}
By further comparing Eq. \eqref{eq: nudeff} with Eq. \eqref{eq: tsc2}, we find the same result as in Eq. \eqref{eq: cntsha} 
at $f_0 = 10^{-6}$ and $\nu_0 = 327$ MHz. 
Therefore, the dependence of $f$ on $\nu$ is given by (Eq. \eqref{eq: defres})
\begin{equation}\label{eq: frelaw}
   f = 10^{-6} \Big(\frac{\nu}{327 \text{MHz}}\Big)^{1.4}.
\end{equation}

Fig. \ref{fig: dataset} is the $\tau_\text{sc}$ vs. DM plot taken from 
\citet{Kri15}.
For comparison, we overplot $\tau_\text{sc, low DM}$ (Eq. \eqref{eq: tsc1}) at low DMs 
and $\tau_\text{sc, high DM}$ (Eq. \eqref{eq: tsc2}) at high DMs, 
which are indeed a good approximation of the fitted $\tau_\text{sc}$-DM relation. 
The slope of the $\tau_\text{sc}$-DM relation flattens at the high-DM end, which comes from the Gaussian distribution of density fluctuations 
on scales smaller than $l_0$. This scattering regime will be discussed in Section \ref{ssec: gaud}.
The equalization of $\tau_\text{sc, low DM}$ (Eq. \eqref{eq: tsc1}) and $\tau_\text{sc, high DM}$ (Eq. \eqref{eq: tsc2}) 
corresponds to the transition between different scattering regimes, with the turnover $\tau_\text{sc}$ and DM, 
\begin{equation}\label{eq: tran}
      \tau_\text{sc,tr} =3.5 \times10^{-3} ~ \text{ms} , ~~
      \text{DM}_\text{tr} =22.7  ~\text{pc cm}^{-3}.
\end{equation}
For nearby pulsars with $\text{DM}<\text{DM}_\text{tr}$, the probability of sightlines to intersect the sparse, discrete density concentrations 
associated with the short-wave-dominated density spectrum is considerably low. 
As a result, the observed scattering is insignificant and 
mainly contributed by the ubiquitous Kolmogorov turbulence for both high- and low-latitude pulsars.  
In contrast, for more distant and low-latitude sources with DM$>\text{DM}_\text{tr}$,
sufficient small-scale density structures are encountered along the propagation path, such that the supersonic turbulence arising in the inner Galaxy 
can manifest itself and dominate the scattering effect. Therefore, 
the resulting scaling of $ \tau_\text{sc}$ with DM is shaped by the short-wave-dominated density spectrum. 
It is necessary to point out that instead of the complete form of the fit (Eq. \eqref{eq: comfm}, the thin solid line in Fig. \ref{fig: dataset}), 
we adopt its asymptotic forms at low- and high-DM limits (Eq. \eqref{eq: tsc1} and \eqref{eq: tsc2}, the thick solid line in Fig. \ref{fig: dataset}) 
for our analysis. As a result, 
the transition between different scattering regimes is sharp. 
In reality, the transition is smoother. However, such a transition is limited to a very short range of DMs, so that the difference between the broken power-law approximation and the smooth-transition model is marginally small (Fig. \ref{fig: dataset}).

Different scattering regimes originate from different turbulence properties. 
As mentioned above, the Kolmogorov turbulence in the WIM has a large driving scale on the order of $100$ pc, 
whereas the supersonic turbulence with a short-wave-dominated density spectrum 
in the inner Galactic plane has an outer scale on the order of a parsec
\citep{Hav08, Malk10}. 
This distinction is also reported in interstellar scattering observations, which imply an outer scale of $\sim 200$ pc
for insignificantly scattered sources in the local ISM and high-latitude active galactic nuclei
\citep{Fr99}, 
but a much smaller outer scale for heavily scattered sources 
(e.g., Sgr A$^*$, NGC6334B, Cyg X-3, see \citealt{NE2} and references therein).
This shows that the scattering model established by involving two types of turbulence is self-consistent.

A two-component model for electron density turbulence including a background widely distributed turbulence and 
occasional discrete plasma structures has been introduced in early investigations on scattering of pulsar radiation 
\citep{Cor85, Lam00, Cor16}.
The density discontinuities discussed in 
\citet{Lam00}
were described by a density spectrum with the spectral slope $\beta = 4$. 
Due to the unclear physical origin and lack of direct evidence from either numerical simulations or observations for 
this special model of density spectrum, the scenario is excluded from our consideration. 
Instead, we adopt a short-wave-dominated density spectrum ($\beta<3$), which is motivated physically and based on both numerical studies 
and observational facts (see Introduction). 
Also, it satisfactorily explains the scaling relation between $\tau_\text{sc,tr}$ and DM for highly scattered pulsars. 
As another difference, the scattering clumps with abrupt density change discussed in these works
are associated with H\uppercase\expandafter{\romannumeral2} regions or supernova shocks. 
When scattering is attributed to discrete dense clumps of a typical size $d$, pulse broadening time is
\citep{Sch68},
\begin{equation}
\begin{aligned}
  \tau_\text{sc}  &= \frac{D^2 r_e^2 \lambda^4}{4\pi c}   \frac{f (\delta n_e(d))^2}{d} \\
                  &= \frac{ r_e^2 c^3}{4\pi  } f \Big(\frac{\delta n_e(d)}{n_e}\Big)^2   d^{-1} \text{DM}^2 \nu^{-4},
\end{aligned}
\end{equation}
which can be evaluated at $\nu = 327$ MHz as 
\begin{equation}
\begin{aligned}
     \tau_\text{sc}  =  4.6\times10^{-9} 
  &     \Big(\frac{f}{10^{-5}}\Big)  \Big(\frac{n_e}{0.01 \text{cm}^{-3}}\Big)^{-2} 
         \Big(\frac{\delta n_e(d)}{1 \text{cm}^{-3}}\Big)^2  \\
  &    \Big(\frac{d}{1 \text{pc}}\Big)^{-1}   \Big(\frac{\text{DM}}{\text{pc cm}^{-3}}\Big)^{2} 
        \Big(\frac{\nu}{327 \text{MHz}}\Big)^{-4} \text{ms},
\end{aligned}
\end{equation}
where the normalization parameters pertain to H\uppercase\expandafter{\romannumeral2} regions 
\citep{HS13}.
By comparing with $\tau_\text{sc, high DM}$ in Eq. \eqref{eq: tsc2}, we find that 
the single-scale clumps of excess electron density fail to produce
the enhanced scattering strength for individual pulsars at high DMs, 
and lead to a DM scaling incompatible with the general observational result. 
In this work, the clumped density structure applied for interpreting the enhancement of scattering 
does not have a single intrinsic length scale but results from a short-wave-dominated power-law density distribution, 
with the relevant density variation and spatial scale much smaller than those of H\uppercase\expandafter{\romannumeral2} regions,
and the DM scaling index dependent on the spectral index of density fluctuations.

The above comparison with the scattering measurements of pulsars not only testifies 
our analytical model for the distribution of interstellar density fluctuations, but also 
allows inferences about the properties of the Kolmogorov turbulence on large scales, as well as much finer
density structures generated by the supersonic turbulence on small scales.

\subsection{Determination of scattering regimes}

\subsubsection{Scattering regimes dominated by the Kolmogorov and supersonic turbulence ($r_\text{diff}>l_0$)}

By formally comparing the analytically derived $\tau_\text{sc}$ as a function of DM with the fit to scattering observations, 
we obtain the typical turbulence parameters appropriate to the interstellar density fluctuations. 
Substituting Eq. \eqref{eq: satconste} into Eq. \eqref{eq: modste}, we find that 
the representative scalings of $\tau_\text{sc}$ with DM and $\nu$ for Galactic pulsars is 
\begin{equation}\label{eq: sufkol}
 \tau_\text{sc,K} = 3.5\times10^5  \Big(\frac{\text{DM}}{\text{pc cm}^{-3}}\Big)^{2.2}  \Big(\frac{\nu}{\text{MHz}}\Big)^{-4.4}  \text{ms}
\end{equation}
in the scattering regime dominated by the Kolmogorov turbulence in the WIM.

For the scattering regime corresponding to the supersonic turbulence with a short-wave-dominated 
density spectrum in the inner Galactic plane, 
in the case of a constant $f$,
using the result given in Eq. \eqref{eq: cntsha}, Eq. \eqref{eq: modsha} leads to
\begin{equation}\label{eq: sufsha}
 \tau_\text{sc,s,cf} = 6.6\times10^{12} \Big(\frac{\text{DM}}{\text{pc cm}^{-3}}\Big)^{4.2}  \Big(\frac{\nu}{\text{MHz}}\Big)^{-8.4}  \text{ms}.
\end{equation}
The resulting scattering time has a strong dependence on both DM and $\nu$. 
As regards the $\nu$-dependent $f$, provided the parameters determined from the pulse-broadening observations
(Eq. \eqref{eq: cntsha}, \eqref{eq: frelaw}), 
the scattering time formulated by Eq. \eqref{eq: nudeff} gives 
\begin{equation}\label{eq: supffre}
    \tau_\text{sc,s,$\nu$f} = 57.0 \Big(\frac{\text{DM}}{\text{pc cm}^{-3}}\Big)^{4.2}  \Big(\frac{\nu}{\text{MHz}}\Big)^{-4}  \text{ms}.
\end{equation}
In comparison with the Kolmogorov scaling in Eq. \eqref{eq: sufkol}, it shows a steeper trend of $\tau_\text{sc}$ with DM, 
but a flatter slope of the $\tau_\text{sc}$-$\nu$ relation. 
We also point out that 
as $f$ is independent of DM, the difference between Eq. \eqref{eq: sufsha} and \eqref{eq: supffre} is only reflected 
in the $\nu$ scaling, with the DM scaling unaffected.

The relative importance between the distinct scaling relations arising from different turbulence regimes 
varies with both DM and $\nu$.
The equality $\tau_\text{sc,K}=\tau_\text{sc,s}$ yields the critical condition for the transition, 
but notice that the transition between different scattering regimes is smooth in realistic situations 
(see Section \ref{ssec: com}).
Thus, we have 
(Eq. \eqref{eq: sufkol} and \eqref{eq: sufsha}) 
\begin{equation}\label{eq: intecl}
     \Big(\frac{\text{DM}}{\text{pc cm}^{-3}}\Big) \Big(\frac{\nu}{\text{MHz}}\Big)^{-2} = 2.3\times10^{-4}
\end{equation}
with a constant $f$, and (Eq. \eqref{eq: sufkol} and \eqref{eq: supffre})
\begin{equation}\label{eq: tranff}
     \Big(\frac{\text{DM}}{\text{pc cm}^{-3}}\Big)^{5} \Big(\frac{\nu}{\text{MHz}}\Big) = 3.0\times10^9
\end{equation}
with a $\nu$-dependent $f$.
It follows that in both cases, the interstellar scattering of nearby pulsars tends to be governed by the Kolmogorov turbulence, 
and the observed scattering time can be estimated using Eq. \eqref{eq: sufkol}.
Whereas for highly dispersed pulsars, 
low-latitude sight lines with long path lengths through the Galactic plane are mostly affected by the supersonic turbulence.
Quite interestingly, 
under the assumption of a constant $f$, it indicates that the pulsars observed at low frequencies tend to be in the supersonic turbulence-dominated 
scattering regime where the observed $\tau_\text{sc}$ is dictated by Eq. \eqref{eq: sufsha}. 
However, with a $\nu$-dependent $f$ adopted, one instead expects the 
dominance of the supersonic turbulence in scattering toward higher frequencies, where
the scaling relation given by Eq. \eqref{eq: supffre} applies. 
From the observational point of view, 
the two scenarios can be easily distinguished given the scattering measurements over a broad range of frequencies (see the next section).

\begin{figure*}[htbp]
\centering
\subfigure[Constant $f$]{
   \includegraphics[width=8cm]{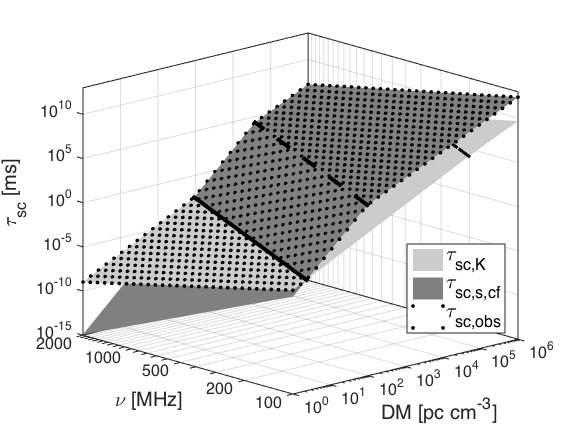} \label{fig: tausca}}
\subfigure[$\nu$-dependent $f$]{
   \includegraphics[width=8cm]{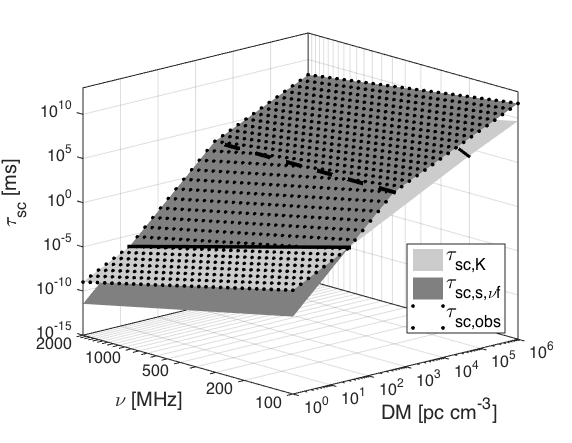} \label{fig: tauscb}}
\caption{ $\tau_\text{sc}$ as a function of both DM and $\nu$.
Light and dark gray planes show the scattering time corresponding to the Kolmogorov ($\tau_\text{sc,K}$, Eq. \eqref{eq: sufkol})
and supersonic turbulence ($\tau_\text{sc,s,cf}$, Eq. \eqref{eq: sufsha} in (a) and $\tau_\text{sc,s,$\nu$f}$, Eq. \eqref{eq: supffre} in (b)), respectively. 
The solid line shows their intersection calculated from Eq. \eqref{eq: intecl} in (a) and Eq. \eqref{eq: tranff} in (b).
Dots denote $\tau_\text{sc,obs}$.
The division between the scattering regimes with $r_\text{diff}>l_0$ and $r_\text{diff}<l_0$ is shown by the 
dashed line for the supersonic turbulence (Eq. \eqref{eq: detreg} in (a) and Eq. \eqref{eq: crsupff} in (b)), 
and by the dash-dotted line for the Kolmogorov turbulence (Eq. \eqref{eq: borkl}). 
$\tau_\text{sc}$ at higher DMs in the regime of $r_\text{diff}<l_0$ is given by  
Eq. \eqref{eq: shinn} in (a) and Eq. \eqref{eq: shfrdf} in (b) for the supersonic turbulence, 
and by Eq. \eqref{eq: kolinn} for the Kolmogorov turbulence.}
\label{fig: tausc}
\end{figure*}

Fig. \ref{fig: tausc} presents the scatter broadening time over a range of DM and $\nu$ for both Kolmogorov and supersonic turbulence.
The observed scattering time $\tau_\text{sc,obs}$ is determined by the maximum between them.
The intersecting line corresponds to the transition between the two scattering regimes dominated by different types of turbulence. 
Besides, we also display the $\tau_\text{sc}$ in the scattering regime with $r_\text{diff}<l_0$, which will be discussed in the next section.

\subsubsection{Scattering regime 
dominated by the Gaussian density distribution ($r_\text{diff}<l_0$)}\label{ssec: gaud}

The above results hold when the inner scale is sufficiently small so that the relation $r_\text{diff} > l_0$ stands, 
but in the case of $r_\text{diff} < l_0$, with the Gaussian tail of the density spectrum (Eq. \eqref{eq: orisf}), 
the density fluctuation at the inner scale is responsible for scattering. 
Then the scaling of $\tau_\text{sc}$ with DM and $\nu$ should be described by 
Eq. \eqref{eq: dissdm}.
By again applying the turbulence parameters given in Eq. \eqref{eq: satconste} and \eqref{eq: cntsha}, and combining 
Eq. \eqref{eq: dissdm} with Eq. \eqref{eq: betpa} and \eqref{eq: denss}, we obtain 
\begin{equation}\label{eq: kolinn}
    \tau_\text{sc,K} = 4.8\times10^5 \Big(\frac{\text{DM}}{\text{pc cm}^{-3}}\Big)^2  \Big(\frac{\nu}{\text{MHz}}\Big)^{-4}  \text{ms}
\end{equation}
for the Kolmogorov density spectrum, and 
\begin{equation}\label{eq: shinn}
     \tau_\text{sc,s,cf} = 2.4\times10^{8} \Big(\frac{\text{DM}}{\text{pc cm}^{-3}}\Big)^2  \Big(\frac{\nu}{\text{MHz}}\Big)^{-4}  \text{ms}
\end{equation}
for the short-wave-dominated density spectrum with a constant $f$. 
Notice that when deriving Eq. \eqref{eq: kolinn}, 
we adopt the same $l_0 \sim 10^7 $ cm of the short-wave-dominated density spectrum for the Kolmogorov spectrum,
which is supported by earlier observations 
\citep{Spa90, Armstrong95, Bha04}.
When the dependence of $f$ on $\nu$ is taken into account, Eq. \eqref{eq: dissdm} at $\beta<3$ is modified as 
\begin{equation}
   \tau_\text{sc,s,$\nu$f} =  \frac{r_e^2 c^3}{4\pi} \mathcal{C}(\beta) g_0 \Big(\frac{\delta n_e }{n_e}\Big)^2 l_0^{-1} \text{DM}^2 
   \nu^{\eta-4}.
\end{equation}
We then use the values of the parameters indicated in Eq. \eqref{eq: cntsha}, \eqref{eq: gamval}, \eqref{eq: frelaw}, and derive from the above equation
\begin{equation}\label{eq: shfrdf}
   \tau_\text{sc,s,$\nu$f} =  8.5\times10^4 \Big(\frac{\text{DM}}{\text{pc cm}^{-3}}\Big)^2  \Big(\frac{\nu}{\text{MHz}}\Big)^{-2.6}  \text{ms}.
\end{equation}
It reveals an even flatter slope of the $\tau_\text{sc}$-$\nu$ relation than that in both Eq. \eqref{eq: kolinn} and \eqref{eq: shinn}.

The criterion for distinguishing between the scattering regimes of $r_\text{diff}>l_0$ and $r_\text{diff}<l_0$ has been presented in Section \ref{sec: scal}.
Given the necessary turbulence parameters (Eq. \eqref{eq: satconste}, \eqref{eq: cntsha}, \eqref{eq: frelaw}), 
Eq. \eqref{eq: parcad} leads to 
\begin{equation}\label{eq: borkl}
   \Big(\frac{\text{DM}}{\text{pc cm}^{-3}}\Big) \Big(\frac{\nu}{\text{MHz}}\Big)^{-2} =  4.9  
\end{equation}
for the Kolmogorov turbulence,
\begin{equation}\label{eq: detreg}
    \Big(\frac{\text{DM}}{\text{pc cm}^{-3}}\Big) \Big(\frac{\nu}{\text{MHz}}\Big)^{-2}  = 9.6\times10^{-3}
\end{equation}
for the short-wave-dominated density spectrum with a constant $f$, and 
\begin{equation}\label{eq: crsupff}
    \Big(\frac{\text{DM}}{\text{pc cm}^{-3}}\Big) \Big(\frac{\nu}{\text{MHz}}\Big)^{-0.625}  = 27.7
\end{equation}
for the short-wave-dominated density spectrum with a $\nu$-dependent $f$.
The above equations for the 
division between the scattering regimes of $r_\text{diff}>l_0$ and $r_\text{diff}<l_0$ 
can be also obtained by equating the scattering time in the two regimes
(i.e., Eq. \eqref{eq: sufkol} and \eqref{eq: kolinn}, Eq \eqref{eq: sufsha} and \eqref{eq: shinn}, Eq. \eqref{eq: supffre} and \eqref{eq: shfrdf}).

\begin{figure*}[htbp]
\centering
\subfigure[Constant $f$]{
   \includegraphics[width=8cm]{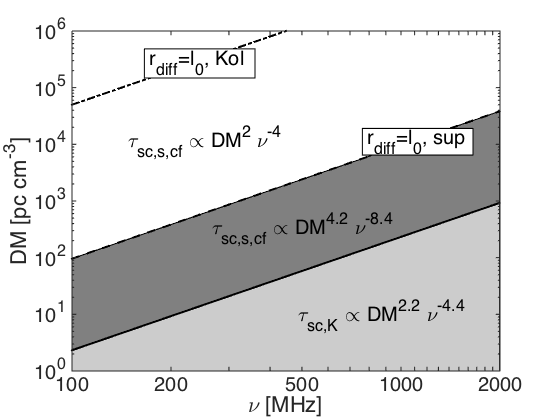}\label{fig: rega}}
\subfigure[$\nu$-dependent $f$]{
   \includegraphics[width=8cm]{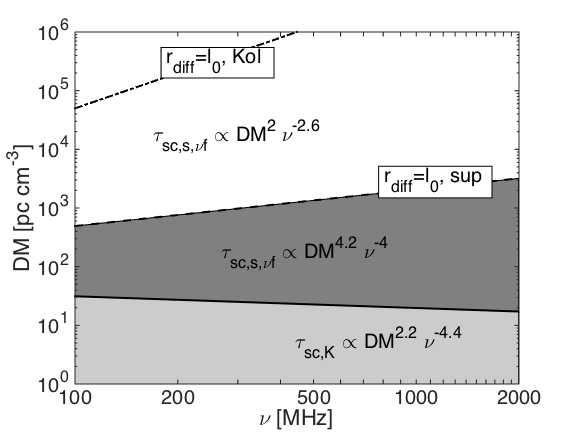}\label{fig: regb}}
\subfigure[Constant $f$, observational data superposed]{
   \includegraphics[width=8cm]{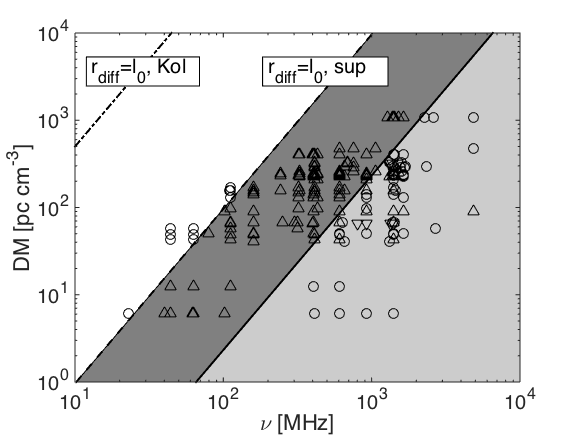}\label{fig: regc}}
\subfigure[$\nu$-dependent $f$, observational data superposed]{
   \includegraphics[width=8cm]{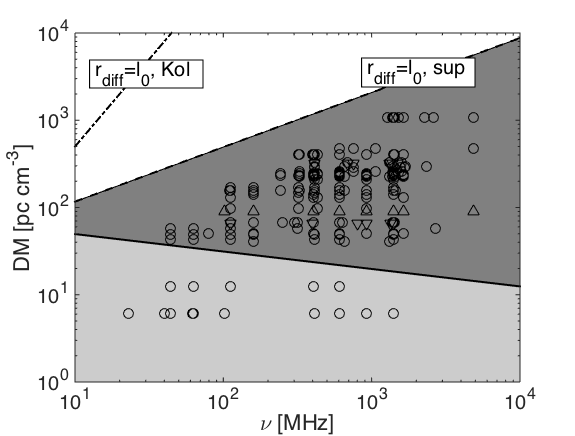}\label{fig: regd}}
\caption{DM and $\nu$ ranges for the scattering regimes dominated by the Kolmogorov and supersonic turbulence with $r_\text{diff}>l_0$ 
(light and dark grey shaded areas), and the scattering regime with $r_\text{diff}<l_0$. 
The transitions between different regimes are indicated by the solid line (Eq. \eqref{eq: intecl} in (a) and Eq. \eqref{eq: tranff} in (b)), the 
dashed line (Eq. \eqref{eq: detreg} in (a) and Eq. \eqref{eq: crsupff} in (b)), and the dash-dotted line (Eq. \eqref{eq: borkl}).
The same symbols are used in (c) and (d), but with the observational data from 
\citet{Lew15}
superposed on the predicted scattering regimes. 
Circles, upward-pointing triangles, and downward-pointing triangles indicate our results are comparable to, overestimate, or underestimate the actual 
$\nu$ scaling index according to the observational measurements. }
\label{fig: reg}
\end{figure*}

Figure \ref{fig: reg} presents the parameter space of DM and $\nu$ for the scattering regimes dominated by the 
Kolmogorov and supersonic turbulence at $r_\text{diff}>l_0$, as well as the regime attributed to the 
Gaussian distribution of density fluctuations at $r_\text{diff}<l_0$. 
With regards to the frequency scaling of $\tau_\text{sc}$,
in Fig. \ref{fig: regc} and \ref{fig: regd}, we also display the multifrequency scattering measurements taken from 
\citet{Lew15}, 
where they provided the largest sample of pulsars with multifrequency estimates of pulse broadening to date. 
With some doubtful results excluded (see their table 1), each data point represents a measurement 
at one of the observing frequencies. 
That is, there are multiple data points with the same DM value but different frequencies corresponding to an individual pulsar. 
Since the $\tau_\text{sc}$ measurements suffer from various sources of error, 
e.g., the error estimates listed in table 1 in \citealt{Lew15} 
range from $0.02$ to $0.86$
(see more discussions on other possible sources of errors in \citealt{Bha04,Lew13}),
when comparing the $\nu$ scaling index derived from our analysis with the observationally measured value, 
we consider our result as ``consistency" if their difference is within the range $[-1,1]$, 
an ``overestimation" if the difference is larger than $1$, 
and an ``underestimation" if the difference is smaller than $-1$.
Obviously, by adopting a $\nu$-dependent $f$, we see a good agreement between the model predictions and observational measurements (Fig. \ref{fig: regd}), 
whereas in the case of a constant $f$, all the $\nu$ scaling indices in the scattering regime dominated by the supersonic turbulence 
are overestimated (Fig. \ref{fig: regc}).

The above results demonstrate that the scaling of $\tau_\text{sc}$ with $\nu$ is consistent with the Kolmogorov scaling of turbulence 
over a broad range of $\nu$ when the DM is sufficiently small ($<100$ pc cm$^{-3}$), which confirms   
earlier observational results, e.g., 
\citet{Cor85, Joh98, Loh01,Loh04, Lew13}.
At higher DMs, by introducing a $\nu$-dependent $f$, the resulting $\nu$ scaling of $\tau_\text{sc}$ in the scattering regime dominated by the supersonic 
turbulence can be also reconciled with the observational results, showing a shallower spectral slope than that expected from a Kolmogorov turbulent medium.

There exist other effects on weakening the $\nu$ dependence of $\tau_\text{sc}$. 
The effect of a finite inner scale of the density power spectrum in the scattering regime $r_\text{diff}<l_0$ has been discussed in 
e.g. \citet{Bha04,NE2}).
But we find that unless in the range of very high DMs, most scattering measurements of pulsars are not in the scattering regime with $r_\text{diff}<l_0$
(see Fig. \ref{fig: reg})
and thus this effect due to the finite inner scale is irrelevant.
Besides, 
another plausible explanation as discussed in 
\citet{CoL01}
is that a transversely truncated scattering screen can result in increasing deficit of scattering at lower frequencies. 
which may be potentially taken into account by modifying the $\nu$ dependence of $f$ in our calculations.
This subject warrants more detailed analysis in future work.

We note that all sizable samples of pulsars compiled for the $\tau_\text{sc}$-DM relation analysis in the literature
(e.g., \citealt{Ram97, Loh04, Bha04, Kri15, Cor16})
include subsamples
which were initially measured at different frequencies and are scaled to the same reference frequency to compose the entire dataset. 
The Kolmogorov scaling is commonly employed for this assembly.
Our results provide the physical justification for this approach in the parameter space of low DMs. 
For higher-DM pulsars, 
a shallower scaling than the Kolmogorov one is more appropriate.

\section{Discussion}

The existence of both subsonic to transonic turbulence with a Kolmogorov density spectrum and 
highly supersonic turbulence with a short-wave-dominated density spectrum in the ISM is supported by many independent 
observational facts. 
The significance of the distribution of density fluctuations in the latter case has not been investigated 
in earlier studies on interstellar scattering of pulsars. 
The scattering measurements of the Galactic pulsars turn out to be a very handy and powerful tool to probe the electron density distribution and 
the associated interstellar turbulence properties. 
Notice that a global analysis of the scattering behavior of a large sample of pulsars 
brings forth the space-averaged features of turbulent density. 
Low-latitude LOSs are subject to local variations in turbulence properties and density inhomogeneities toward the inner Galaxy, 
leading to a large scatter in $\tau_\text{sc}$ about the overall $\tau_\text{sc}$-DM relation, as well as in the $\nu$ scaling index
for high-DM pulsars
\citep{Jo98, Cor02, Bha04, Lew13}.

The short-wave-dominated density spectrum arises in supersonic turbulence, which is a common state of the cold and dense media in the inner Galaxy. 
In the case of collapsing clouds, due to the effect of self-gravity, 
the density spectrum can undergo a transition from the turbulence- to gravity-dominated regime toward 
smaller scales, with the 1D spectral slope changing from a negative value to a positive value 
\citep{Bur15}. 
Since the probability for the LOS to intersect with a star-forming region is relatively low compared to the supersonic turbulent media, 
here we did not take this situation into account in our statistical analysis of the interstellar scattering for a large sample of Galactic pulsars,
but the gravity-modified density spectrum 
\citep{Bur15}
can be important for interpreting the scattering measurements of individual pulsars in particular directions toward 
collapsing clouds.

Besides in the Galactic ISM, 
the presence of supersonic turbulence is also expected in the host galaxies of extragalactic radio sources 
which are undergoing active star formation.
The associated short-wave-dominated density spectrum 
results in the scatter broadening of the observed pulse width of e.g., a fast radio burst
\citep{XZf}.

\section{Conclusions}

Under the consideration of the two populations of turbulent density fields
in the diffuse ionized ISM and in the cold and dense ISM phases in the Galactic plane,
we construct a spectral model for the Galactic distribution of electron density fluctuations. 
Our main conclusions are summarized as follows:

\begin{enumerate}[(1)]

\item By comparing with the scattering measurements of pulsars, 
we identify a scattering regime dominated by the Kolmogorov turbulence 
for low-DM pulsars,  
and a more enhanced scattering regime dominated by the supersonic turbulence which is characterized by a short-wave-dominated density spectrum
with the spectral index $\beta \approx 2.6$ (corresponding to $M_s \approx 7$)
for low-latitude and high-DM pulsars.

\item By introducing a $\nu$-dependent filling factor $f$ in the scattering regime dominated by the supersonic turbulence, the spectral model of density 
fluctuations that we constructed can also explain the shallower scaling of $\tau_\text{sc}$ with $\nu$ in comparison with the Kolmogorov scaling.
Despite the small sample of pulsars measured at a few frequencies and considerable uncertainties in $\tau_\text{sc}$ measurements 
due to e.g., dispersion smearing, low signal-to-noise ratio,
this model is supported by the available multifrequency observations of pulsars with relatively large DMs over a broad range of $\nu$.

\item By comparing our analytical model with pulsar observations, 
we obtained the relations that impose observational constraints on 
the fundamental properties of the ISM turbulence. To satisfy these relations, we found plausible values of 
the energy injection scale $L$, electron density fluctuation over the length scale $L$ in the Kolmogorov turbulence (Eq. \eqref{eq: satconste}),
\begin{equation}
 L \sim 100 ~\text{pc}, ~~ \delta n_e \sim 0.04 ~\text{cm}^{-3}, 
\end{equation}
and the characteristic spatial scale, electron density, and volume filling factor of small-scale density irregularities in the supersonic turbulence 
(Eq. \eqref{eq: cntsha}, \eqref{eq: frelaw}),
\begin{equation}
\begin{aligned}
  &  l_0 \sim 10^7 ~\text{cm}, ~~ \delta n_e \sim 0.1 ~\text{cm}^{-3}, \\
  &  f \sim 10^{-6} \Big(\frac{\nu}{327 \text{MHz}}\Big)^{1.4}.
\end{aligned}
\end{equation}

\item 
We provide the parameter space of DM and $\nu$ for different scattering regimes and 
corresponding scalings of $\tau_\text{sc}$ (see Fig. \ref{fig: reg}), 
which can be useful for designing future large-scale and scattering-limited pulsar surveys.

\end{enumerate}

The spectral model for interstellar density fluctuations proposed in this work 
for explaining interstellar scattering measurements as well as probing the interstellar turbulence 
should be further tested and refined 
with a finer frequency sampling of more accurate scatter broadening measurements by using the 
forthcoming data from, e.g., LOFAR   
\citep{van13},
the MWA
\citep{Tin12}, 
the SKA.

\acknowledgments
We thank the anonymous referee for helpful comments.
We thank Dipanjan Mitra for useful discussions. 
This work is partially supported by the National Basic Research Program (973 Program) of China under grant No. 2014CB845800.

\bibliographystyle{apj.bst}
\bibliography{yan}

\end{document}